\documentstyle[aps,preprint]{revtex}
\begin{document}
\tightenlines

\draft
\title{Effective Action and Conformal Phase Transition in
Three--Dimensional ${\rm QED}$}
\author{V. P. Gusynin\cite{email1}$^{1}$,
V. A. Miransky\cite{email2}$^{1}$,
and A. V. Shpagin$^{2}$}

\address{$^{1}$Bogolyubov Institute for Theoretical Physics
\protect\\
252143 Kiev, Ukraine\protect\\
$^{2}$Institute of Physics, 252450 Kiev, Ukraine
\protect\\ }
\maketitle

\begin{abstract}
The effective action for local composite operators in $QED_3$ is considered. 
The effective potential is calculated in leading order in
$1/N_f$ ($N_f$ is the number of fermion flavors) and used to describe the
features of the phase transition at $N_f=N_{\rm cr}$, $3<N_{\rm cr}<5$.
It is shown that this continuous phase transition satisfies the criteria of
the conformal phase transition, considered recently in the literature. In
particular, there is an abrupt change of the spectrum of light excitations
at the critical point, although the phase transition is continuous, and the
structure of the equation for the divergence of the dilatation current is
essentially different in the symmetric and nonsymmetric phases. The
connection of this dynamics with the dynamics
in $QCD_4$ is briefly discussed.
\end{abstract}

\pacs{11.30.Rd, 11.10.Kk, 11.10.Gh, 12.20.Ds}
\section{Introduction}
Noncompact quantum electrodynamics in $2+1$ dimensions ($QED_3$) with
$N_f$ flavors of four-component fermions \cite{pisarsky84,appelquist}
is a gauge theory with rich dynamics, reminiscent of four-dimensional
quantum chromodynamics ($QCD_4$).

Studying the Schwinger-Dyson (SD) equation for the fermion self-energy
in leading order in $1/N_f$ expansion \cite{appnashwij} showed
the existence
of a critical number of fermion flavors $N_{\rm cr}$, $3<N_{\rm cr}<5$,
below which there is the dynamical breakdown
of the flavor $U(2N_f)$ symmetry
and fermions acquire a dynamical mass. These conclusions were confirmed
in the lattice computer simulations of noncompact lattice $QED_3$
\cite{dagotto}.

The presence of a critical $N_{\rm cr}$ in $QED_3$ is intriguing especially
because of a possibility of the existence of an analogous critical $N_f=
N_{\rm cr}$ in $QCD_4$, suggested by both analytical studies
\cite{banks,apptewij,miryam97,chivukula} and lattice
computer simulations of $QCD_4$ \cite{kogut,brown,iwasaki}.

However there is still controversy concerning the existence of a finite 
$N_{\rm cr}$ in $QED_3$: some authors argue that the generation of a fermion 
mass occurs at all values of $N_f$ in $QED_3$ \cite{mike,pisarsky91}. Despite
quite intensive studies of this issue \cite{kondo,GHR},
the problem is still 
open. The main difficulty is that at present there is no systematic approach 
to studying nonperturbative strong coupling dynamics in gauge theories. A 
possibility to get insight into the phase transition at $N_f=N_{\rm cr}$ 
might be a low energy effective action approach, which has been very 
successful in studying the phase transition at finite temperatures in
$QCD_4$ \cite{piswil}. Such an approach is useful when there are few relevant 
(massless) degrees of freedom at the critical point. For example, at 
$T=T_{\rm cr}$ in $QCD_4$, the relevant degrees of freedom are $N_f^2$ 
(or $N_f^2-1$)
pseudoscalar and $N_f^2$ scalar mesons: because of a non-zero temperature,
and therefore non-zero Matsubara frequences for quarks,
all other excitations are effectively massive at
$T=T_{\rm cr}$. As a result, the effective theory is given by a three 
dimensional {\it renormalizable} lagrangian with the complex matrix field 
$\phi_i^j$, $i,j=1,2,\dots,N_f$, describing $2N_f^2$ scalar and 
pseudoscalar mesons \cite{piswil}.
This allows one to reduce the problem to a known
universality class of three dimensional models.

Should such an approach (reducing the infrared dynamics to that of a known
universality class) work in the case of the phase transition at
$N_f=N_{\rm cr}$ in $QED_3$? We believe that the answer is "no". The point
is that there are too many relevant degrees of freedom at $N_f=N_{\rm cr}$
in $QED_3$. Besides a composite field $\varphi$, describing $2N_f^2$
Nambu-Goldstone (NG) bosons (corresponding to the spontaneous breakdown 
$U(2N_f)\rightarrow U(N_f)\times U(N_f)$)
and their $2N_f^2-1$ flavor partners
(for details, see below), there are massless fermions and massless photon at 
$N_f=N_{\rm cr}$. Therefore, unlike the temperature phase transition in 
$QCD_4$, the effective action includes even more degrees of freedom
than the initial lagrangian. This in turn implies that the effective action 
for the field $\varphi$, that can be obtained when {\it light} fermion and 
photon fields are  integrated out, is necessary non-local and
non-renormalizable. In other words, the universality class of the phase 
transition in $QED_3$ is different from those described by
renormalizable ( or even local ) lagrangians 
for the field $\varphi$.

Still, it would be interesting to find the
effective action for $\varphi$ in
$QED_3$, describing the phase digram
of the theory. In this paper, we make a
step in realizing this program and calculate the effective potential in 
leading order in $1/N_f$ expansion. Our main goal is to figure out how the 
structure of the potential reflects some peculiar properties of the phase 
transition at $N_f=N_{\rm cr}$ in $QED_3$ pointed out in 
Refs.\cite{miryam97,apptewij95}. It gives
an example of the 
conformal phase transition \cite{miryam97}, the conception
which generalizes 
the Berezinsky-Kosterlitz-Thouless (BKT)
phase transition  \cite{BKT} ( taking place 
in two dimensions ) to higher dimensions.

We stress that our aim is much more modest than to prove the existence of a
finite $N_{\rm cr}$ in $QED_3$. We will just derive the effective potential,
accepting, following Ref.\cite{appnashwij}, that the
leading order in $1/N_f$ is a  reliable approximation
in the theory. Then we will see that the
structure of this potential indeed corresponds to the conformal phase 
transition. In particular, it will be shown that unlike the usual,
$\sigma$-model-like, continuous phase transition, there is an
abrupt change of the spectrum of light excitations at the critical
point in $QED_3$. Also, the realization of the conformal symmetry
is very different in the symmetric and non-symmetric phases.

\section{The Model. General Properties.}
The lagrangian density for massless $QED_3$ is
\begin{equation}
{\cal L}=-{1\over 4}F_{\mu\nu}^2+\bar\psi i{\hat D}\psi,
\label{lagrangian}
\end{equation}
where $D_{\mu}=\partial_\mu-ieA_\mu$,
${\hat D}=\gamma^{\mu}D_{\mu}$,
and four-component fermion fields carry 
the flavor index $i=1,2,\dots,N_f$. The three $4\times4$ $\gamma$-matrices 
can be taken to be
\begin{equation}
\gamma^0= \left(\begin{array}{cc} \sigma_3 & 0 \\ 0& -
\sigma_3\end{array}\right),\quad
\gamma^1= \left(\begin{array}{cc} i\sigma_1&0\\0&-
i\sigma_1\end{array}\right),\quad
\gamma^2= \left(\begin{array}{cc}i\sigma_2&0\\0&-i\sigma_2\end{array}\right).
\label{eq:2}
\end{equation}
Recall that $(\sigma_3,i\sigma_1,i\sigma_2)$ and   $(-\sigma_3,-i\sigma_1,-
i\sigma_2)$  make two inequivalent representations
of the Clifford algebra in $2+1$ dimensions.

There are two matrices
\begin{equation}
\gamma^3= i\left(\begin{array}{cc} 0& 1 \\ 1& 0\end{array}\right),\quad
\gamma^5= i\left(\begin{array}{cc} 0&1\\-1&0\end{array}\right),
\end{equation}
that anticommute with  $\gamma^0,\gamma^1$ and $\gamma^2$. Therefore for
each four-component spinor, there is a global $U(2)$ symmetry with
generators
\begin{equation}
I,\quad {1\over i}\gamma^3,\quad \gamma^5,\quad{\rm and}\quad {1\over 2}
[\gamma^3,\gamma^5],
\end{equation}
and the full symmetry is then $U(2N_f)$.

The lagrangian density with a mass term $m\bar\psi\psi$ is invariant under
parity transformations defined as
\begin{equation}
P:\qquad \psi(x^0,x^1,x^2)\rightarrow {1\over i}\gamma^3\gamma^1\psi(x^0,
-x^1,x^2).
\end{equation}
The generation of a parity-invariant dynamical mass for fermions leads
to the spontaneous breakdown of the $U(2N_f)$ down to $U(N_f)\times U(N_f)$
with the generators
\begin{equation}
\frac{\lambda^\alpha}{2},\quad 
\frac{\lambda^\alpha}{2}{1\over2}[\gamma^3,\gamma^5],
\end{equation}
$\alpha=0,1,\dots, N_f^2-1$. The corresponding $2N_f^2$ NG bosons are:
\begin{equation}
s^\alpha\sim \bar\psi\frac{\lambda^\alpha}{2}\gamma^5\psi
\end{equation}
($N_f^2$ scalars), and
\begin{equation}
p^\alpha\sim \bar\psi\frac{\lambda^\alpha}{2}\gamma^3\psi
\end{equation}
($N_f^2$ pseudoscalars). There are also their $2N_f^2$ massive flavor
partners:
\begin{equation}
S^\alpha\sim \bar\psi\frac{\lambda^\alpha}{2}\psi
\end{equation}
($N_f^2$ scalars), and
\begin{equation}
P^\alpha\sim \bar\psi\frac{\lambda^\alpha}{2}{1\over 2}[\gamma^3,\gamma^5]
\psi
\end{equation}
($N_f^2$ pseudoscalars).
Notice that these $4N_f^2$ bosons can be described by the
hermitian  matrix field
\begin{equation}
\phi_j^i=\psi^i\psi^{\dag}_j
\end{equation}
which can be decomposed into a traceless part and its trace
defined as:
\begin{equation}
\varphi\equiv \phi-{1\over 8N_f}\gamma^0[\gamma^5,\gamma^3]\chi,\quad \chi
\equiv tr
({1\over 2}\gamma^0 [\gamma^5,\gamma^3]\phi)\sim P^0.
\end{equation}
The field $\varphi$ is assigned to the adjoint representation of $SU(2N_f)$,
and $\chi$ is a singlet. As to the vacuum group $U(N_f)\times U(N_f)$, the
$2N_f^2$ NG bosons are assigned to the representation $(N_f,N_f^*)\oplus
(N_f^*,N_f)$, and the $2N_f^2$ massive bosons are assigned to the
representation $(N_f
\times N_f^*,1)\oplus (1,N_f\times N_f^*)=
(N_f^2 - 1, 1)\oplus (1, N_f^2 - 1)\oplus 2(1,1)$ of its maximal
semi-simple subgroup $SU(N_f)\times SU(N_f)$.
One of the
two  singlets
$(1,1)$ corresponds to the pseudoscalar 
$\chi= {1\over 2}\bar\psi[\gamma^3,\gamma^5]\psi\sim P^0$,
and another to the
scalar $\sigma\equiv \bar\psi\psi\sim S^0$.
The vacuum expectation value $\langle 
0|\sigma|0\rangle$ is an order parameter describing
the spontaneous breakdown $U(2N_f)\rightarrow U(N_f)\times U(N_f)$.

Recall that besides the generation of a parity-invariant mass, corresponding
to $\sigma_c\equiv\langle 0|\sigma|0\rangle\neq0$, there might be the
generation of a $U(2N_f)$-invariant mass corresponding to $\chi_c\equiv
\langle0|\chi|0\rangle\neq0$, thus violating parity. Here we accept arguments
of Ref.\cite{appbokawij} in the support of a solution with a parity-invariant
vacuum.

In leading order in $1/N_f$, in Landau gauge, the SD equation for the fermion
propagator $G(p)=(A(p)\hat p-\Sigma(p))^{-1}$ is reduced to the equations
\cite{appelquist,appnashwij}:
\begin{eqnarray}
& &\hspace{4cm}A(p)=1,\nonumber\\
& &\Sigma(p)=m_0+\frac{\alpha}{\pi^2N_fp}\left[\int\limits_0^p
dk\frac{k\Sigma(k)}{k^2+\Sigma^2(k)}\frac{k}{p+{\alpha\over8}}+
\int\limits_p^\Lambda
dk\frac{k\Sigma(k)}{k^2+\Sigma^2(k)}\frac{p}{k+{\alpha\over8}}\right],
\label{SDeqs}
\end{eqnarray}
where $\alpha\equiv e^2N_f$, $\Lambda$ is an ultraviolet cutoff,
and, for generality, the bare fermion mass
$m_0$ is introduced.

Notice that the fermion mass function is
neglected in the vacuum polarization in this approximation.
The reliability of this approximation has been
studied in Refs.\cite{kondo,GHR}. In particular, in Ref.\cite{GHR},
to keep $A(p)$=1, a non-local gauge was used. It was shown then
that including the mass function in the vacuum polarization
does not lead to qualitative changes of the results of Ref.\cite
{appnashwij}: the scaling law for the dynamical mass function
near the critical point has the same form ( see Eq.(\ref{dynmass})
below), with a somewhat different value of $N_{\rm cr}$:
$N_{\rm cr}=4.3$
instead $N_{\rm cr}=3.2$. Moreover, this new value agrees
with the result
of the second paper of Ref.\cite{appnashwij}, incorporating
the next-to-leading corrections in $1/N_f$ to the SD equation in
Landau gauge.

That study suggests that neglecting a fermion mass
function in the
vacuum polarization is a reasonable approximation for a small
($\Sigma_0 << \alpha$) fermion mass, and this condition is
fulfilled for $N_f$ close
to the critical point $N_{\rm cr}$.
Therefore including the mass function in the vacuum
polarization
seems do not change the qualitative features of the phase
transition at $N_f=N_{\rm cr}$. Since the main aim of this paper
is to give a qualitative description of this phase transition,
the use of this approximation seems appropriate.

We will return to equation (\ref{SDeqs}) in the next section.

\section{Effective Potential for Local Composite Fields in $QED_3$}
We will consider the effective action (generating functional for proper
vertices) of the local composite field
$\varphi$ in $QED_3$. Actually we will
calculate only the effective potential, i.e. the part of the action without
derivatives of $\varphi$, though we will also describe properties of
other terms in the effective action which are necessary for
establishing the origin of the phase transition.

In the derivation of the effective potential, we will closely follow Ref.
\cite{miransky}, where the effective potential for local composite fields in
quenched $QED_4$ was derived.

The effective action is defined in the standard way. First, one introduces
a generating functional for Green's functions of the field $\varphi$:
\begin{equation}
Z(J)=\exp(iW(J))=\int d\eta\exp\left[i\int d^3x\left({\cal L}(x)+ tr
\left[J(x)
\varphi(x)\right]\right)\right],
\label{genfunc}
\end{equation}
where the $\eta$ integration is functional, $J(x)$ is the source for
$\varphi(x)$, and ${\cal L}(x)$ is the lagrangian density (\ref{lagrangian})
(the symbol $\eta$ represents all the fundamental
fields (fermion and photon ones) of the model).

The effective action for the field $\varphi$ is a Legendre transform of the
functional $W(J)$:
\begin{equation}
\Gamma(\varphi_c)=W(J)-\int d^3x tr \left[J(x)\varphi_c(x)\right],
\label{Gamma}
\end{equation}
where $\varphi_c(x)\equiv\langle0|\varphi(x)|0\rangle$.
>From Eqs.(\ref{genfunc})
and (\ref{Gamma}) one finds that the following relations are satisfied:
\begin{eqnarray}
\frac{\delta W}{\delta J(x)}&=&\varphi_c(x),\\
\frac{\delta \Gamma}{\delta\varphi_c(x)}&=&-J(x).
\label{source}
\end{eqnarray}
The effective action $\Gamma$ can be expanded in powers of derivatives of
the field $\phi_c$:
\begin{equation}
\Gamma(\varphi_c)=\int d^3x\left[-V(\varphi_c)+
{1\over2}Z(\varphi_c) tr (\partial_\mu
\varphi_c\partial^\mu\varphi_c) +\dots\right],
\end{equation}
where $V(\varphi_c)$ is the efective potential.

The calculation of the effective potential is reduced to finding the
Legendre transform of the functional $W(J)$ with the source $J$ independent
of coordinates $x$. The field $\varphi_c$ is a matrix in the flavor space
describing $4N_f^2-1$ boson fields.

Let us first consider the case when a constant source is introduced for the
composite field $\sigma=\bar\psi\psi$: ${\cal L}\rightarrow{\cal L}+J\bar
\psi\psi$. Then the source term in $Z(J)$
has the form of a bare mass term, with $m_0\equiv-J$.
Eq.(\ref{source}) now becomes
\begin{equation}
\frac{dV}{d\sigma_c}=-m_{0}.
\end{equation}
Therefore
\begin{equation}
V=-\int\limits^{\sigma_c}m_0{(x)}dx,
\label{eq:pot}
\end{equation}
where $m_0(\sigma_c)$ defines the dependence of $m_0=-J$ on the
condensate $\sigma_c=\langle0|\sigma|0\rangle=\langle0|\bar\psi\psi|0
\rangle$, which is an order parameter of the spontaneous breakdown
$U(2N_f)\rightarrow U(N_f)\times U(N_f)$. The function $m_0(\sigma_c)$
can be defined from the SD equations (\ref{SDeqs}).

The effective potential $V$ is calculated in the Appendix. It is:
\begin{eqnarray}
V&=&\frac{A^2N_f^2\Sigma_0^3}{32}\left[-{7\over3}+\frac{3}{4\nu^2}-
{1\over \nu}\sin{2\theta}-\left(1+\frac{3}{4\nu^2}\right)\cos{2\theta}
\right]\nonumber\\
&+&\frac{A^2N_f\Lambda\Sigma_0^3({9\over 4}+\nu^2)}{4\pi^2\nu^2
\Lambda_{\rm np}}\sin^2(\theta+\nu\delta_1),
\label{effpot}
\end{eqnarray}
where $\nu=\sqrt{\lambda-1/4},\quad\lambda=8/\pi^2N_f$, $\Lambda$ is an
ultraviolet cutoff, $\Lambda_{\rm np}\equiv\alpha/8$  is an ultraviolet
cutoff for nonperturbative dynamics, like $\Lambda_\chi\equiv4\pi F_\pi
\sim 1$Gev in $QCD_4$ \cite{manohar,mirbook}
(for details, see the Appendix),
$\Sigma_0\equiv\Sigma(p^2)|_{p=0}$ is an infrared
fermion mass parameter,
\begin{eqnarray}
\theta=\nu\log\frac{\Lambda_{\rm np}e^{\delta_0}}{\Sigma_0},\quad
\delta_0=3\log2+{\pi\over2}-2,\quad 
\delta_1={1\over\nu}\arctan\frac{2\nu}{3},\nonumber
\end{eqnarray}
and
\begin{eqnarray}
A=\frac{\sqrt\pi}{2(1+{8\over\pi^2N_f})}\bigg|\frac{\Gamma(1+i\nu)}
{\Gamma({1\over4}+{i\nu\over2})\Gamma({5\over4}+{i\nu\over2})}
\bigg|.\nonumber
\end{eqnarray}
Notice that expression (\ref{effpot}) for $V$ is valid at $\Sigma_0<<
\Lambda_{\rm np}=\alpha/8$.

The dependence of $\Sigma_0$ on the condensate $\sigma_c$ is defined from
the equation:
\begin{equation}
\sigma_c=\langle0|\bar\psi\psi|0\rangle=-\frac{2N_f\Lambda m_0}{\pi^2}+
\sigma_c^{\rm np},
\label{sigmac}
\end{equation}
where the nonperturbative part of the condensate is
\begin{equation}
\sigma_c^{\rm np}=\frac{N_f^2}{4}\Lambda^2_{\rm np}\frac{d\Sigma}{dp}
\bigg|_{p=\Lambda_{\rm np}}\simeq\frac{N_f^2}{4}\Lambda^2_{\rm np}\Sigma_0
\left(\frac{d}{dp}F({1\over4}-{i\nu\over2},{1\over4}+{i\nu\over2},{3\over2};
-\frac{p^2}{\Sigma_0^2})\right)\bigg|_{p=\Lambda_{\rm np}},
\label{sigmanp}
\end{equation}
$F$ is a hypergeometric function (for details, see the Appendix). The
parameter
$m_0$ is expressed through $\Sigma_0$ as
\begin{equation}
m_0=\frac{A\Sigma_0^{3/2}\sqrt{9/4+\nu^2}}{2\nu\Lambda^{1/2}_{\rm np}}
\sin(\nu\log\frac{\Lambda}{\Sigma_0}+\nu(\delta_0+\delta_1)).
\label{baremass}
\end{equation}
Eqs.(\ref{effpot}),(\ref{sigmac}),(\ref{sigmanp}), and (\ref{baremass})
define $V$ as an implicit, and rather complicated, function of $\sigma_c$.
However, the phase diagram of the theory can be established by studying
$V$ as a function of the fermion mass $\Sigma_0$. It is convenient to
rewrite Eq.(\ref{effpot}) as
\begin{eqnarray}
V&=&\frac{A^2N_f^2\Sigma_0^3}{32}\left[-{7\over3}+\frac{3}{4\nu^2}+
\left(1-\frac{3}{4\nu^2}\right)\cos({2\theta}+2\nu\delta_1)-
\frac{2}{\nu}\sin({2\theta}
+2\nu\delta_1)\right]\nonumber\\
&+&\frac{A^2N_f\Lambda\Sigma_0^3({9/4}+\nu^2)}{4\pi^2\nu^2
\Lambda_{\rm np}}\sin^2(\theta+\nu\delta_1).
\label{effpotential}
\end{eqnarray}
Then the gap equation is:
\begin{eqnarray}
\frac{dV}{d\Sigma_0}&=&\frac{A^2N_f^2\Sigma_0^2}{16}\left\{\left[\left(
-7+\frac{9}{4\nu^2}\right)\sin(\theta+\nu\delta_1)+\left(2\nu-
\frac{15}{2\nu}\right)\cos(\theta+\nu\delta_1)\right]\right.\nonumber\\
&+&\left.\frac{4\Lambda({9/4}+\nu^2)}{N_f\pi^2\nu^2\Lambda_{\rm np}}
\left[3\sin(\theta+\nu\delta_1)-2\nu\cos(\theta+\nu\delta_1)\right]\right\}
\sin(\theta+\nu\delta_1)=0.
\end{eqnarray}
This equation yields the following solutions:
\begin{equation}
\Sigma_0^{(n)}=\Lambda_{\rm np}\exp\left(-\frac{\pi n}{\nu}+\delta_0
+\delta_1\right),\quad n=1,2,\dots ,
\label{minima}
\end{equation}
\begin{equation}
\Sigma_0^{(n)}=\Lambda_{\rm np}\exp\left(-\frac{\pi n}{\nu}+\delta_0
+\delta_1-{2\over 3}\right),\quad n=1,2,\dots\quad .
\label{maxima}
\end{equation}
One can check that while all the solutions (\ref{minima}) correspond to 
minima of $V$, solutions (\ref{maxima}) correspond to maxima of the
potential. Actually, only the global minimum, corresponding to $n=1$,
defines the stable vacuum \cite{mirbook}. Therefore the dynamical mass is
\begin{equation}
m_{\rm dyn}\equiv\bar\Sigma_0=\Lambda_{\rm np}\exp\left(-\frac{\pi}{\nu}
+\delta_0+\delta_1\right).
\label{dynmass}
\end{equation}
As it has to be, it coincides with $\bar\Sigma_0$ of Ref.\cite{appnashwij},
derived from the SD equation.

Few comments are in order:
\begin{enumerate}
\item Since expressions (\ref{effpot})
and (\ref{effpotential}) for $V$ are 
valid only if $\Sigma_0<<\Lambda_{\rm np}$, the solution (\ref{dynmass})
exists when $0<\nu<<1$, i.e.
when $0<{8\over \pi^2N_f}-{1\over4}<<1$.
Therefore, in this
approximation, there is a critical value $N_f=N_{\rm cr}={32\over 
\pi^2}\simeq 3.24$, and expression (\ref{dynmass}) is valid in the near-
critical, scaling, region.
\item The last term in expression (\ref{effpot}) for $V$ is connected with a
perturbative contribution, i.e. with ultraviolet dynamics at $p>>\Lambda_{
\rm np}=\alpha/8$, where the dimensionless running coupling constant is
weak (see Eq.(\ref{runcoupling}) below).
This term occurs because of the presence of
a source $J=-m_0\neq0$ outside
the extrema of $V$.
On the other hand, since this term is proportional to
$\sin^2(\theta+\nu\delta_1)$, the {\it dynamical}
mass (\ref{dynmass}), defined at the minimum,
is independent of the perturbative contribution.
Notice also that while this contribution is proportional to $N_f$, the
nonperturbative contribution in $V$ is proportional to $N_f^2$.

The perturbative term is divergent. It is connected with the point
that we calculate the effective action for local {\it composite}
operators, which is a generating functional for proper vertices
(Green's functions) of these operators. It is known that
the standard renormalizations do not remove divergences from
such Green's functions (for example, see Sec.12.15
in the book \cite{mirbook}).
These divergences are connected with perturbative short-range
fluctuations, having nothing with infrared non-perturbative
dynamics of bound states. The receipt of dealing with them in
studying bound states
(used, for example, in the case of QCD sum rules, dealing with
Green's functions of colorless composite operators) is known:
one has just to subtract them. Then we are led to the
non-perturbative effective potential:
\begin{equation}
V_{\rm np}=\frac{A^2N_f^2\Sigma_0^3}{32}\left[-{7\over3}+\frac{3}{4\nu^2}-
{1\over \nu}\sin{2\theta}-\left(1+\frac{3}{4\nu^2}\right)\cos{2\theta}
\right].
\label{npeffpot}
\end{equation}
It contains only contributions from the nonperturbative
dynamics and is independent of the cutoff $\Lambda$. Notice
that in the symmetric phase, with $N_f>N_{\rm cr}$, the parameters
$\theta$ and $\nu$ in Eq. (\ref{npeffpot}) are imaginary.

Henceforth the cutoff $\Lambda$ is put equal infinity and we will
consider the region of the nonperturbative dynamics
with momenta $p<\Lambda_{\rm np}$.
\item We have derived the effective potential
for the composite operator 
$\sigma=\bar\psi\psi$. In principle, one can derive the effective potential
for other $4N_f^2-2$ boson fields, described by the composite field
$\varphi$,
by introducing additional $4N_f^2-2$ sources independent of $x$
The
calculations of this potential at $N_f>1$ are involved, and the expression
for $V$ should be very cumbersome. We have not succeeded in getting it.
Fortunately, the effective action for $\sigma_c$, with all other fields 
taken equal zero, is sufficient for our purposes. The reason is the
following.
On the one hand, since
$\sigma_c$ is the only order parameter of the spontaneous breakdown
$U(2N_f)\rightarrow U(N_f)\times U(N_f)$, the effective action for $\sigma_c$
defines the mass of $\sigma$ particle both in symmetric and nonsymmetric
phases. On the other hand, as was already pointed out
in the previous section,
in the symmetric phase, all the $4N_f^2-1$ bosons connected
with the composite field $\varphi$ are assigned to the same (adjoint)
irreducible representation of the $SU(2N_f)$. This point will be enough
for us for proving that the phase transition in $QED_3$ is indeed
conformal.
\end{enumerate}

\section{Conformal Phase Transition in $QED_3$}

The characteristic feature of the phase transition at $N_f=N_{\rm cr}$
in $QED_3$ is that the scaling function
\begin{equation}
f(N_f)=\exp\left(-\frac{\pi}{\nu}+\delta_0+\delta_1\right),\quad \nu=
{1\over2}\sqrt{\frac{N_{\rm cr}}{N_f}-1}
\end{equation}
($N_{\rm cr}={32\over\pi^2}$ in this approximation), has an essential
singularity at $N_f=N_{\rm cr}$. In particular, considering the scaling
function $f(z)$ as an analytic function of the complex variable $z=N_f$,
one finds that while $\lim_{z\to z_{\rm cr}}f(z)=0$ as $z$ goes to
$z_{\rm cr}$ from the side of nonsymmetric phase ($N_f<N_{\rm cr}$),
$\lim_{z\to z_{\rm cr}}f(z)\neq 0$ as $z\to z_{\rm cr}$ from the
side of the symmetric phase with $N_f>N_{\rm cr}$.
This feature implies that the continuous 
phase transition in $QED_3$ satisfies the criteria
of the conformal phase transition
(CPT) introduced in Ref.\cite{miryam97}. ( Notice that since
$N_f$ appears analytically in the path integral of the theory,
one can give a nonperturbative meaning to the theory with noninteger
$N_f$ ).

As shown in Ref.\cite{miryam97}, the CPT is characterized by the following 
two general features: a) unlike the usual ($\sigma$-model-like) continuous
phase transition, there is an abrupt change of the spectrum of 
light excitations at the critical point,
though the CPT is a continuous phase transition; b) the realization of the
conformal symmetry is very different in the symmetric and nonsymmetric
phases: in particular, the equation for the divergence of the dilatation
current, $\partial^\mu D_\mu$, has essentially different structures in
those two phases.

In this section, we will discuss how these two features are reflected
in the effective action in $QED_3$.

$QED_3$ is a super-renormalizable theory where the coupling constant is
dimensional. However, as was pointed already in Ref.\cite{appelquist},
in the infrared region, with $p<<\Lambda_{\rm np}=\alpha/8$, the conformal
symmetry is a good symmetry in the symmetric phase with $N_f>N_{\rm cr}$
and $m_{\rm dyn}=0$ (at least in leading order in $1/N_f$). The point is
that in that region, the dimensional coupling constant $\alpha$ drops
out from the SD equations for Green's functions: this, in particular, can
be seen on the example of the SD equation for the mass function $\Sigma(p)$
(see Eq.(\ref{A4}) in the Appendix). This point is also connected with the
fact that the dimensionless running coupling constant
\begin{equation}
\bar\alpha(p)\equiv\frac{\alpha}{8p(1+\Pi(p))}
\label{runcoupling}
\end{equation}
($\Pi(p)$ is a polarization operator) has an infrared stable fixed point
$\bar\alpha=1$ in the symmetric phase \cite{appelquist}.

Of course, in the nonsymmetric phase, because of the generation of the
fermion dynamical mass, the conformal symmetry is broken.

As it was already pointed out in the previous section,
the parameter $\Lambda_{\rm np}$
is an effective ultraviolet cutoff of the nonperturbative dynamics.
Since, in leading order in $1/N_f$, the polarization operator
$\Pi(p)=\alpha/8p=\Lambda_{\rm np}/p $
in the symmetric phase \cite{appelquist}, the
running coupling (\ref{runcoupling}) is $\bar\alpha=1$ at all values of
$p$ as $\Lambda_{\rm np}\to\infty$.
Therefore the dynamics is conformal invariant in that phase
in this limit.

The physical meaning of this limit is the following. The near-critical
dynamics around $N_f=N_{\rm cr}$ implies that,
for such values of $N_f$, masses of light
excitations have to be much less than the parameter $\Lambda_{\rm np}$.
We are interested in the low energy
dynamics of these excitations, when their energies are much less than
$\Lambda_{\rm np}$. The question is whether there are such light
excitations described by the composite field $\sigma$ (and, in general,
by the field $\varphi$). If they are, then the field $\Sigma_0$,
describing their fluctuations around the minimum of the
effective potential $V_{\rm np}$ ( see Eq. (\ref{npeffpot})),
has also to be much less than $\Lambda_{\rm np}$.
This in turn implies that the potential $V_{\rm np}$,
describing the nonperturbative dynamics,
has to be much less than
$\Lambda_{\rm np}^3$.

This leads us to studying the auxiliary "continuum"
limit $\Lambda_{\rm np}$ goes to infinity
with $\Sigma_0$ and $V_{\rm np}$ being finite
(the critical dynamics).
As usual for critical dynamics
\cite{mirbook}, such a limit can be interpreted as a renormalization
of the parameter $N_f$. In other words, one has to find such a
dependence of $N_f$ on $\Lambda_{\rm np}$ that as
$\Lambda_{\rm np}$ goes to infinity, the effective potential
$V_{\rm np}$ would be finite for finite $\Sigma_0$.

It is not difficult to show from Eqs. (\ref{dynmass})
and (\ref{npeffpot})
that in the nonsymmetric phase the
non-perturbative effective potential is in this limit:
\begin{equation}
V_{\rm np}=A^2N_{\rm cr}^2\Sigma_0^3\left(-{1\over24}+{1\over8}
\log\frac{\Sigma_0}{\bar
\Sigma_0}+{3\over64}\log^2\frac{\Sigma_0}{\bar\Sigma_0}\right).
\label{finnpeffpot}
\end{equation}
As it has to be, $N_f$, defined from Eq. (\ref{dynmass}),
is fixed as $\Lambda_{\rm np}\to\infty$: $N_f=N_{\rm cr}$.

But what is the form of the potential
at $N_f=N_{\rm cr}$ from the side of the symmetric
phase in this limit?
\footnote{Since $N_f$ is dimensionless, it is actually a
function of the ratio $\Lambda_{\rm np}/\mu$, where $\mu$ is a
renormalization group parameter of the dimension of mass. It can
be introduced in different ways. For example, one condition
might be that the potential $V_{\rm np}$ is equal to
$(\mu)^3$ at $\Sigma_0=\mu$.
In the end of the calculations,
if this limit exists, $\mu$ can be
expressed through such physical parameters as the mass of
$\sigma$ boson (in both symmetric and asymmetric phases) or
the dynamical fermion mass in the asymmetric phase
(see equation (\ref{finnpeffpot})).}
It is easy to
show that in the symmetric phase, at finite $\Sigma_0$, the potential
$V_{\rm np}$ (\ref{npeffpot}) diverges as $\Lambda_{\rm np}$
goes to infinity (on technical side, it is connected with the point
that $\sin{2\theta}$ and $\cos{2\theta}$ in the potential become
hyperbolic functions for
$N_f>N_{\rm cr}$).
In particular, for $N_f$ close to $N_{\rm cr}$, the potential is:
\begin{equation}
V_{\rm np}\rightarrow\frac{A^2N^2_{\rm cr}\Sigma_0^3}{32}\left[-{10\over3}-
2\log\frac{\Lambda_{\rm np}e^{\delta_0}}{\Sigma_0}+{3\over2}
\log^2\frac{\Lambda_{\rm np}e^{\delta_0}}{\Sigma_0}
\right]\rightarrow\infty
\label{continuumV}
\end{equation}
as $\Lambda_{\rm np}\to\infty$.

On the other hand, the kinetic term and terms with a larger
number of derivatives (the structure of which is the same in the
symmetric and nonsymmetric phases) have to be finite in this
limit. Indeed, the most severe ultraviolet divergences always occur
in an effective potential. Since the potential $V_{\rm np}$
(\ref{continuumV})
diverges only logarithmically, all other terms in the effective action
are finite.

This situation implies that, in the symmetric phase, there are
no $\it{light}$ particles described
by the potential $V_{\rm np}$. If they exist, they
are heavy (with $M\sim\Lambda_{\rm np}$) and
therefore decouple from infrared dynamics.

In the present case, $\sigma$
boson is such a particle. However, since, as was already
pointed out in Sec.2,
in the symmetric phase, all the $4N_f^2-1$ particles
described by the field $\varphi$ are in the same
(adjoint) representation of the $SU(2N_f)$, all of them decouple there.
Therefore, in the limit $\Lambda_{\rm np}\to\infty$,
the symmetric phase in massless $QED_3$ is
a Coulomb-like, conformal-invariant
phase, describing interactions between massless fermions and photons.

On the other hand, the non-perturbative
effective potential, and the whole effective action,
for this composite field is finite as $\Lambda_{\rm np}\to\infty$
in the nonsymmetric phase
(see Eq.(\ref{finnpeffpot})). This reflects the point
that in that phase, besides photons and fermions, there are other light
(with a mass $M<<\Lambda_{\rm np}$) particles: $2N_f^2$ massless
composite NG bosons and  $2N_f^2-1$ massive composite bosons with
a mass $M\sim m_{\rm dyn}\equiv
\bar\Sigma_0<<\Lambda_{\rm np}$ (in the scaling, near-critical, region).
Moreover, the conformal symmetry is explicitly broken in the continuum
limit in the nonsymmetric phase. Indeed, as follows from 
Eq.(\ref{finnpeffpot}),
\begin{equation}
\langle0|\partial^\mu D_\mu|0\rangle=\langle0|\theta^\mu_\mu|0\rangle=
4V_{\rm np}(\Sigma_0)\bigg|_{\Sigma_0=\bar\Sigma_0}=
-\frac{A^2N_f^2}{6}\bar
\Sigma_0^3\neq0,
\end{equation}
where $\theta^\mu_\nu$ is the energy-momentum tensor. Therefore
there is a conformal anomaly in this phase, and the vacuum
expectation value $\langle0|\theta^\mu_\mu|0\rangle$ plays here the
same role as the gluon condensate in $QCD_4$.

Thus, in leading order in $1/N_f$, the phase transition in $QED_3$
possesses all the characteristic features of the conformal phase
transition. In particular, in agreement with the general
conclusion of Ref. \cite{miryam97}, there is no Ginzburg-Landau
like effective action describing the near-critical dynamics in
$\it{both}$ symmetric and broken phases.

It is instructive to compare the dynamics in $QED_3$ with the
dynamics in $QED_3$ with a Chern-Simons (CS) term. As is known
\cite{km,hong}, the CS term enforces the phase transition to be
first order. Therefore an abrupt change of the spectrum of
light excitations at the critical point is natural in that case
\cite{hong}. In $QED_3$ without the CS term, there is still an
abrupt change of the spectrum, though the phase transition
is continuous.

It is amazing how closely the dynamics in $QED_3$ resembles the dynamics
in quenched $QED_4$ \cite{mirbook,FGMS}. The scaling law for $m_{\rm dyn}$
in $QED_4$ has the form (\ref{dynmass}) with $\Lambda_{\rm np}$ replaced
by cutoff $\Lambda$ and $N_{\rm cr}/N_f$ replaced by $\alpha^{(4)}/
\alpha^{(4)}_{\rm cr}$, where the dimensionless critical coupling
$\alpha^{(4)}_{\rm cr}\sim 1$. The effective potential (\ref{npeffpot})
also closely resembles the effective potential calculated in Refs.
\cite{miransky,miryam97}. This implies that long-range dynamics, provided
by strong Coulomb-like ($\sim1/r$) forces, are essentially the same in these 
two models.

Recently, the existence of the conformal phase transition (CPT) in
quenched $QED_4$ has been confirmed by calculating the spectrum of
composites directly from Schwinger-Dyson equations for Green's
functions of local composite operators \cite{vm}. It would be worth
considering such an approach in $QED_3$.

The CPT is a conception extending the BKT phase transition \cite{BKT}
(taking place in two dimensions) to higher dimensions. To see this in
$QED_3$, it is convenient to consider the continuum limit $\Lambda_{\rm np}
\rightarrow\infty$ with $m_{\rm dyn}=\bar\Sigma_0$ being fixed as a
renormalization of the parameter $N_f^{-1}$. Then, the scaling law
(\ref{dynmass}) implies the following $\beta$ function at $N_f<N_{\rm cr}$:
\begin{equation}
\beta(N_f^{-1})=
\frac{\partial N_f^{-1}}{\partial\log \Lambda_{\rm np}}
=-{\pi\over32}\left(
\frac{N_{\rm cr}}{N_f}-1\right)^{3/2}.
\label{betha}
\end{equation}
Clearly, this function has an ultraviolet stable fixed point at
$N_f=N_{\rm cr}$.
On the other hand, since at $N_f>N_{\rm cr}$, where $m_{\rm dyn}=0$, the
infrared dynamics with $p<<\Lambda_{\rm np}=\alpha/8$ is essentially
independent of $\Lambda_{\rm np}$, the $\beta$ function is identically
zero. Therefore there is a line of fixed points at $N_f>N_{\rm cr}$.

Such a renormalization group (RG) structure coincides with that
corresponding to the BKT phase transition.

Of course, real $QED_3$ has a fixed integer $N_f$ and a finite
$\Lambda_{\rm np}=\alpha/8$. Still, the $\beta$-function (\ref{betha})
is useful in
the description of its dynamics. Let us consider a RG invariant quantity
$X$. Then:
\begin{equation}
\frac{dX}{d\log\Lambda_{\rm np}}=\frac{\partial X}{\partial\log
\Lambda_{\rm np}}+\beta(N_f^{-1})\frac{\partial X}{\partial N_f^{-1}}=0.
\end{equation}
This relation demonstrates that the $\beta$ function $\beta(N_f^{-1})$
defines how $X$ depends on the coupling constant $\alpha=8\Lambda_{\rm np}$.

Thus the CPT yields extension of the BKT phase transition to higher
dimensions: the crucial property of those phase transitions
is the presence of an essential singularity in the mass (energy) gap
at the critical point. However, there is an essential difference
between
the realization of the CPT in two and higher dimensions.
While there cannot be spontaneous breakdown of any continuous symmetry
in two dimensions \cite{MWC}, there is no such a restriction in higher
dimensions. In particular, there is a genuine spontaneous flavor
symmetry breaking in $QED_3$.

\section{Conclusion}
In this paper we have analyzed the structure of the effective action in
$QED_3$ in leading order in $1/N_f$ and showed that it reflects
the existence of the conformal phase transition at $N_f=N_{\rm cr}$. How
much of this picture survives beyond the $1/N_f$ expansion
is still an open issue,
though results of the lattice computer simulations \cite{dagotto} are
encouraging.

The existence of such a phase transition, which is an extension of the
BKT phase transition to higher dimensions ($2+1$, in this case), is
interesting in itself. Also, the CPT in $QED_3$
may be connected with nonperturbative
dynamics in condensed matter, in particular, with
dynamics of a non-Fermi liquid \cite{dorey,aitchison}.

The most interesting issue is the connection of the phase transition in
$QED_3$ with a possibility of the existence of an analogous phase
transition in a $(3+1)$-dimensional $SU(N_c)$ gauge theory
\cite{apptewij,miryam97}. In $QED_3$ the critical number $N_f=N_{\rm cr}$
separates two phases with very different dynamics. At $N_f>N_{\rm cr}$
there is a Coulomb phase describing interactions of massless photons and
fermions; at $N_f<N_{\rm cr}$ the rich dynamics with both spontaneous
flavor symmetry breaking and confinement is realized. These
two nonperturbative phenomena
are intimately connected in this model:
at $N_f<N_{\rm cr}$ massive fermions decouple
from infrared dynamics, thus leading to a potential growing with
a distance, and therefore to confinement.
As was already mentioned above, strong Coulomb-like forces play important
role in providing the CPT in $QED_3$.

Is a similar picture realized in a $(3+1)$-dimensional $SU(N_c)$ gauge
theory? What is the interplay between the dynamics provided by strong
Coulomb-like forces and those connected with topologically nontrivial
fluctuations, like instantons, monopoles, etc.? $QED_3$, with its rich
dynamics, is a fruitful laboratory for studying those complicated issues.

\section{Acknowledgments}

The work of V.P.G. is supported by Swiss National Science
Foundation grant CEEC/NIS/96-98/7 IP 051219 and by Foundation of
Fundamental Researches of Ministry of Sciences of Ukraine under
grant No. 2.5.1/003 . V.A.M. acknowledges useful discussions with
R. Pisarski. He is grateful to Direccio General de la Recerca de
la Generalitat de Catalunya for financing his visit to the Group
de Fisica Teorica of the Univ. Autonoma de Barselona where part
of this research was carried out.
He is also grateful to the CSSM/NITP hosted by the University of
Adelaide for their hospitality during the Workshop on Nonperturbative
Methods in Quantum Field Theory, during which this paper
was completed.

\appendix
\section{Derivation of the effective potential}
In this Appendix we derive expression (\ref{effpot}) for the effective
potential $V$.

First, we need to recall some properties of the solution of the SD equation
(\ref{SDeqs}). This equation will be used to extract the dependance of
the parameter $m_0$ on the condensate $\sigma_c$.

Differentiating the second of Eqs.(\ref{SDeqs}) with respect
to $p$, one finds that $\Sigma(p)$ satisfies the differential equation
\begin{equation}
{d\over dp}\left[p^2\frac{(p+\alpha/8)^2}{2p+\alpha/8}\frac{d\Sigma}
{dp}\right]=-\frac{\alpha}{\pi^2N_f}\frac{p^2\Sigma(p)}{p^2+\Sigma^2(p)}
\label{A1}
\end{equation}
with infrared (IRBC) and ultraviolet (UVBC) boundary conditions:
\begin{eqnarray}
& &p^2\frac{d\Sigma}{dp}\bigg|_{p=0}=0\qquad ({\rm IRBC}),\\
\label{A2}
& &\left[\frac{p(p+\alpha/8)}{2p+\alpha/8}\frac{d\Sigma}{dp}+\Sigma(p)
\right]\bigg|_{p=\Lambda}\simeq\left[{\Lambda\over2}\frac{d\Sigma}
{dp}+\Sigma(p)\right]\bigg|_{p=\Lambda}=m_{0}\quad ({\rm UVBC}).
\label{A3}
\end{eqnarray}
Following Ref.\cite{appnashwij}, one can approximate the equation
(\ref{A1}) by writing it in two regions as follows:
\begin{equation}
{d\over dp}\left[p^2\frac{d\Sigma}{dp}\right]=-
\frac{8}{\pi^2N_f}\frac{p^2\Sigma}{p^2+\Sigma^2_0}, \quad
p<\Lambda_{\rm np}\equiv{\alpha\over8},
\label{A4}
\end{equation}
\begin{equation}
{d\over dp}\left[p^3\frac{d\Sigma}{dp}\right]=-
\frac{2\alpha}{\pi^2N_f}\Sigma, \quad
p>\Lambda_{\rm np},
\label{A5}
\end{equation}
where, in Eq.(\ref{A4}) , $\Sigma^2(p)$ in the denominator is replaced
by $\Sigma^2_0\equiv\Sigma(p)|_{p=0}$; in Eq.(\ref{A5}), at
$p>\Lambda_{\rm np}$, $\Sigma^2(p)$  in the denominator is neglected.
This is justified if $\Sigma^2(p)<<\Lambda_{\rm np}$ : as we will see,
it is indeed correct in the near-critical region.

The parameter $\Lambda_{\rm np}=\alpha/8$ is an effective ultraviolet cutoff 
for nonperturbative dynamics in the model: at $m_0=0$, the dynamical mass 
function rapidly decreases at $p\geq\Lambda_{\rm np}$ (see below). This
parameter plays here the same role as the parameter $\Lambda_\chi=4\pi
F_\pi\sim1$Gev in $QCD_4$ \cite{manohar,mirbook}.

Taking into account IRBC (\ref{A2}), the solution to Eq.(\ref{A4}) is
expressed through a hypergeometric function:
\begin{equation}
\Sigma(p)=\Sigma_0F\left({1\over 4}-{i\nu\over2},{1\over4}+{i\nu\over2},
{3\over2};-\frac{p^2}{\Sigma^2_0}\right),
\label{A6}
\end{equation}
where $\nu=\sqrt{\frac{8}{\pi^2N_f}-{1\over4}}$. At $p>>\Sigma_0$, its
asymptotics is \cite{gradshtein}
\begin{equation}
\Sigma(p)=A_0\frac{\Sigma_0^{3/2}}{\nu\sqrt p}\sin\left(\nu\log\frac{p}
{\Sigma_0}+\nu\delta_0\right),
\label{A7}
\end{equation}
where
\begin{equation}
A_0={\sqrt\pi\over2}\bigg|\frac{\Gamma(1+i\nu)}{\Gamma({1\over4}+{i\nu
\over2})\Gamma({5\over4}+{i\nu\over2})}\bigg|
\label{A8}
\end{equation}
and
\begin{equation}
\delta_0={1\over\nu}\arg \left[\frac{\Gamma(1+i\nu)}{\Gamma({1\over4}+{i\nu
\over2})\Gamma({5\over4}+{i\nu\over2})}\right]
\label{A9}
\end{equation}
(notice that $\delta_0=3\log2+{\pi\over2}-2$ as $\nu\to0$).

Solutions of Eq.(\ref{A5}) can in principle be expressed through Bessel
functions, but, for our purposes, it is sufficient to display them as a
series in $x=p/\alpha$:
\begin{equation}
\Sigma=x^s\left(C_0+\frac{C_1}{x}+\cdots\right)\Rightarrow\Sigma=C_0
\left(1+\frac{2}{\pi^2N_f}{\alpha\over p}\right)+C_2\frac{\alpha^2}{p^2}
+\cdots .
\label{A10}
\end{equation}
Then we find from Eq.(\ref{A3}):
\begin{equation}
m_{0}=C_0\left(1+\frac{1}{\pi^2N_f}{\alpha\over\Lambda}\right).
\label{A11}
\end{equation}
Notice that Eqs.(\ref{A10}) and (\ref{A11}) imply that at $m_{0}=0$
the dynamical mass function rapidly (as ${1\over p^2}$) decreases with
increasing $p$ in the region with $p\geq \Lambda_{\rm np}$.
Therefore, as was
already stated above, $\Lambda_{\rm np}\sim\alpha$ is indeed an ultraviolet
cutoff for nonperturbative dynamics.

Matching now solutions of equations (\ref{A4}) and (\ref{A5}) at
$p=\Lambda_{\rm np}$ (i.e. equating the values of the functions and their
derivatives), one gets:
\begin{equation}
\Sigma(p)\bigg|_{p=\Lambda_{\rm np}}=C_0\left(1+\frac{2}{\pi^2N_f}
{\alpha\over\Lambda_{\rm np}}\right)+C_2\frac{\alpha^2}
{\Lambda_{\rm np}^2},
\label{A12}
\end{equation}
\begin{equation}
\frac{d\Sigma}{dp}\bigg|_{p=\Lambda_{\rm np}}=-\frac{2C_0}{\pi^2N_f}
{\alpha\over\Lambda_{\rm np}^2}-2C_2\frac{\alpha^2}{\Lambda_{\rm np}^3}.
\label{A13}
\end{equation}
Eqs.(\ref{A11}),(\ref{A12}), and (\ref{A13}) imply
\begin{equation}
m_0=\frac{1}{1+\frac{8}{\pi^2N_f}}\left(\Sigma+{\Lambda_{\rm np}\over 2}
\frac{d\Sigma}{dp}\right)\bigg|_{p=\Lambda_{\rm np}}
\label{A14}
\end{equation}
(here we neglect a term of order $\alpha\over\Lambda$).
Then, using this equation and Eq.(\ref{A7}), we arrive at the equation
\begin{equation}
m_0=\frac{A\Sigma_0^{3/2}}{2\nu\sqrt{\Lambda_{\rm np}}}\sqrt{{9\over4}+
\nu^2}\sin\left(\nu\log\frac{\Lambda_{\rm np}}{\Sigma_0}+\nu(\delta_0+
\delta_1)\right),
\label{A15}
\end{equation}
where
\begin{eqnarray*}
A=\frac{A_0}{1+\frac{8}{\pi^2N_f}},\quad \delta_1={1\over\nu}\arctan\frac
{2\nu}{3}.
\end{eqnarray*}
At $m_0=0$ we find solutions for the dynamical mass (compare with Eq.
(\ref{minima})):
\begin{equation}
\Sigma_0^{(n)}=\Lambda_{\rm np}\exp\left(-\frac{\pi n}{\nu}+\delta_0
+\delta_1\right),\quad n=1,2,\dots .
\label{A16}
\end{equation}
Eq.(\ref{eq:pot}) implies that $V$ as a function of $\Sigma_0$ is:
\begin{equation}
V(\Sigma_0)=-\int\limits^{\Sigma_0}m_0(x)\frac{d\sigma_c}{dx}dx.
\label{A17}
\end{equation}
The function $m_0(\Sigma_0)$ is given by Eq.(\ref{A15}). Let us determine
the function $\frac{d\sigma_c}{dx}$. The condensate $\sigma_c$ is:
\begin{eqnarray}
\sigma_c&=&\langle0|\bar\psi(x)\psi(x)|0\rangle=-{\rm tr}
\langle0|T\psi(x)\bar\psi(y)|0\rangle\bigg|_{y\to x} =-i{\rm tr}G(x,x)
\nonumber\\
&=&-\frac{i}{(2\pi)^3}{\rm tr}\int d^3p\frac{1}{\hat p-\Sigma(p)}=
-\frac{2N_f}{\pi^2}\int\limits_0^{\Lambda}
dp\frac{p^2\Sigma(p)}{p^2+\Sigma^2(p)}.
\label{A18}
\end{eqnarray}
The nonperturbative part of the condensate corresponds to the integration
region with $p<\Lambda_{\rm np}$:
\begin{equation}
\sigma_c^{\rm np}=-\frac{2N_f}{\pi^2}\int\limits_0^{\Lambda_{\rm np}}dp
\frac{p^2\Sigma(p)}{p^2+\Sigma^2(p)}=\frac{N_f^2}{4}
\Lambda_{\rm np}^2\frac
{d\Sigma}{dp}\bigg|_{p=\Lambda_{\rm np}}.
\label{A19}
\end{equation}
Here the last equality follows directly from equation (\ref{A4}).

Because, as follows from Eqs.(\ref{A10}) and (\ref{A11}), 
$\Sigma(p\to\infty)=m_0$, the perturbative part of the condensate is
$\sigma_c^{\rm pt}=-\frac{2N_f}{\pi^2}m_0\Lambda$. This and equation
(\ref{A19}) lead to expression (\ref{sigmac}) for the condensate 
$\sigma_c=\sigma_c^{\rm np}+\sigma_c^{\rm pt}$.

Eqs.(\ref{sigmac}),(\ref{A15}) and (\ref{A17}) lead to expression 
(\ref{effpot}) for the effective potential. Notice that the expression
for the potential in the symmetric phase, with
$N_f>N_{\rm cr}$, is an analytic continuation of the potential in
the broken phase, with $N_f<N_{\rm cr}$.


\begin{references}

\bibitem[*]{email1}
            E-mail: {\tt vgusynin@gluk.apc.org}
\bibitem[\dagger]{email2}
            E-mail: {\tt miransky@phys.uconn.edu}


\bibitem{pisarsky84}R.~D.~Pisarski, Phys. Rev. D{\bf29}, 2423 (1984).

\bibitem{appelquist}T.~Appelquist, M.~Bowick, D.~Karabali, and
L.~C.~R.~Wijewardhana, Phys. Rev. D{\bf33}, 3704 (1986).

\bibitem{appnashwij}T.~Appelquist, D.~Nash, and
L.~C.~R.~Wijewardhana, Phys. Rev. Lett. {\bf60}, 2575 (1988);
D.~Nash, {\it ibid.} {\bf62}, 3024 (1989).

\bibitem{dagotto} E.~Dagotto, J.~Kogut, and A.~Koci\'c, Nucl. Phys.
B{\bf334}, 279 (1990); S.~Hands and J.~B.~Kogut, {\it ibid.} B{\bf335},
455 (1990).

\bibitem{banks} T.~Banks and A.~Zaks, Nucl. Phys. B{\bf196}, 189 (1982).

\bibitem{apptewij}T.~Appelquist, J.~Terning, and
L.~C.~R.~Wijewardhana, Phys. Rev. Lett. {\bf77}, 1214 (1996).

\bibitem{miryam97} V.~A.~Miransky and K.~Yamawaki, Phys. Rev. D{\bf55},
5051 (1997).

\bibitem{chivukula}R.~S.~Chivukula, Phys. Rev D{\bf55}, 5238 (1997).

\bibitem{kogut}J.~B.~Kogut and D.~R.~Sinclair, Nucl.Phys. B{\bf295}, 465
(1988).

\bibitem{brown}F.~R.~Brown, H.~Chen, N.~H.~Christ, Z.~Dong, R.~D.~Mawhinney,
W.~Shafer, and A.~Vaccarina, Phys. Rev. D{\bf46}, 5655 (1992).

\bibitem{iwasaki}Y.~Iwasaki, K.~Kanaya, S.~Sakai, and T.~Yoshi\'e, Phys. Rev.
Lett. {\bf69}, 21 (1992).

\bibitem{mike}M.~R.~Pennington and D.~Walsh, Phys. Lett. B{\bf253}, 246 
(1991); D.~C.~Curtis, M.~R.~Pennington, and D.~Walsh, {\it ibid.} {\bf295},
313 (1992).

\bibitem{pisarsky91}R.~D.~Pisarski, Phys. Rev. D{\bf44}, 1866 (1991).

\bibitem{kondo}K.-I.~Kondo and H.~Nakatani, Mod. Phys. Lett. A{\bf5}, 407
(1990); C.~J.~Burden and J.~Praschifka, and C.~D.~Roberts, Phys. Rev. 
D{\bf46}, 2695 (1992); P.~Maris, {\it ibid.} {\bf52}, 6087 (1995);
P.~Maris, {\it ibid.} {\bf54}, 4049 (1996); I.~J.~R.~Aitchison,
N.~E.~Mavromatos, and D.~McNeill, Phys. Lett. B{\bf402}, 154 (1997).

\bibitem{GHR}V.~P.~Gusynin, A.~H.~Hams,
 and M.~Reenders, Phys. Rev. D{\bf53}, 2227 (1996);

\bibitem{piswil}R.~D.~Pisarski and F.~Wilczek, Phys. Rev. D{\bf29}, 338 
(1984).

\bibitem{apptewij95}T.~Appelquist, J.~Terning, and
L.~C.~R.~Wijewardhana, Phys. Rev. Lett. {\bf75}, 2081 (1995).

\bibitem{BKT}V.~L.~Berezinskii, Sov. Phys. JETP{\bf32}, 493 (1970);
J.~M.~Kosterlitz and D.~J.~Thouless, J. Phys. C{\bf6}, 1181 (1973).

\bibitem{appbokawij}T.~Appelquist, M.~Bowick, D.~Karabali, and
L.~C.~R.~Wijewardhana, Phys. Rev. D{\bf33}, 3774 (1986).

\bibitem{miransky}V.~A.~Miransky, Int. J. Mod. Phys. A{\bf8}, 135 (1993).

\bibitem{manohar}A.~Manohar and H.~Georgi, Nucl. Phys. B{\bf234}, 189 (1984);
H.~Georgi, Phys. Lett. B{\bf298}, 187 (1993).

\bibitem{mirbook}V.~A.~Miransky, Dynamical Symmetry Breaking in Quantum
Field Theories (World Scientific, Singapore, 1993).

\bibitem{km}K.-I. Kondo and P. Maris, Phys. Rev. Lett. {\bf74}, 18 (1995).

\bibitem{hong}D. K. Hong, Phys. Rev. D{\bf57}, 1313 (1998).

\bibitem{FGMS} P.~I.~Fomin, V.~P.~Gusynin, V.~A.~Miransky, and Yu.~A.~
Sitenko, Riv. Nuovo Cimento {\bf6}, 1 (1983).

\bibitem{vm} V.~P.~Gusynin and M.~Reenders, Phys. Rev. D{\bf57},
6356 (1998)

\bibitem{MWC} N. D. Mermin and H. Wagner, Phys. Rev. Lett. {\bf17},
1133 (1966); S. Coleman, Commun. Math. Phys. {\bf31}, 259 (1973).

\bibitem{dorey} N. Dorey and N. E. Mavromatos, Phys. Lett. B{\bf250},
107 (1990); A. Kovner and B. Rosenstein, Phys. Rev. B{\bf42},
4748 (1990).

\bibitem{aitchison} I.~J.~R.~Aitchison and N.~E.~Mavromatos, Phys. Rev.
B{\bf53}, 9321 (1996); K.Farakos and N.E. Mavromatos, Int. J. Mod. Phys.
B{\bf12}, 809 (1998).

\bibitem{gradshtein}I.~S.~Gradshtein and I.~M.~Ryzhik, Tables of Integrals,
Series and Products (Academic Press, Orlando, 1980).

\end{references}
\end{document}